\documentclass[aps,prd,preprint,nofootinbib,11pt]{revtex4}
\usepackage{amsfonts}
\usepackage{mathrsfs}
\usepackage{graphicx}
\usepackage{amsmath}
\usepackage{amssymb}
\usepackage{subfigure}
\usepackage{epsfig}
\usepackage{graphicx}
\usepackage{ulem}
\usepackage{color}
\newcommand{\bqa}{\begin{eqnarray}}
\newcommand{\eqa}{\end{eqnarray}}
\newcommand{\beq}{\begin{equation}}
\newcommand{\eeq}{\end{equation}}
\allowdisplaybreaks[1]
\graphicspath{{fig/}{dia/}} \DeclareGraphicsExtensions{.eps}

\begin{document}
\title{\Large Mass spectrum of the $\Omega\bar{\Omega}$ states\\[7mm]}

\author{Bing-Dong Wan$^{1,2}$\footnote{wanbd@lnnu.edu.cn}, Jun-Hao Zhang$^{1,2}$, and Yan Zhang$^{1,2}$\vspace{+3pt}}

\affiliation{$^1$Department of Physics, Liaoning Normal University, Dalian 116029, China\\
$^2$ Center for Theoretical and Experimental High Energy Physics, Liaoning Normal University, Dalian 116029, China}

\author{~\\~\\}

\begin{abstract}
\vspace{0.3cm}
In this study, we investigate the mass spectrum of the $\Omega\bar{\Omega}$ states with quantum numbers $J^{PC}=0^{-+}$, $1^{--}$, $0^{++}$, and $1^{++}$ within the framework of QCD sum rules. Employing suitably constructed interpolating currents, the analyses are carried out with the operator product expansion (OPE) including condensate contributions up to dimension $12$. Our results indicate the existence of four possible baryonium states with masses $m_{0^{-+}}=(3.22\pm0.07)$ GeV, $m_{1^{--}}=(3.28\pm0.08)$ GeV, $m_{0^{++}}=(3.46\pm0.09)$ GeV, and $m_{1^{++}}=(3.54\pm0.11)$ GeV. For the $0^{-+}$ and $1^{--}$ states, the predicted masses lie below the corresponding dibaryon thresholds, suggesting possible bound-state configurations. In contrast, the $0^{++}$ and $1^{++}$ states are found above the respective thresholds, implying resonance-like behavior. Potential decay channels for these baryonium candidates are discussed, with emphasis on those accessible to current experimental facilities such as BESIII, Belle II, and LHCb.
\end{abstract}
 \maketitle
\newpage

\section{Introduction}
Phenomenological investigations of exotic hadronic states provide valuable opportunities to probe the internal structure of hadrons and the underlying mechanisms of confinement in quantum chromodynamics (QCD). The dynamic interplay between experimental discoveries and theoretical developments has significantly advanced our understanding of hadron spectroscopy, offering deeper insights into the fundamental properties of the strong interaction. In recent years, numerous charmonium- and bottomonium-like $XYZ$ states have been observed in various experiments~\cite{Choi:2003ue, Aubert:2005rm, Belle:2011aa, Ablikim:2013mio, Liu:2013dau}, enriching the landscape of hadronic matter and revealing novel manifestations of QCD dynamics.

Motivated by the experimental confirmation of tetraquark and pentaquark states, it is both natural and timely to extend the search for exotic hadronic matter to include possible hexaquark configurations. A paradigmatic example is the deuteron, a bound state of a proton and a neutron, formed in the early Universe, whose stability underpins the synthesis of heavier elements. As an experimentally well-established dibaryon molecular state with quantum numbers $J^P=1^+$ and a binding energy $E_B=2.225$ MeV~\cite{Weinberg:1962hj}, the deuteron serves as a benchmark for understanding nuclear binding mechanisms. The strong nuclear force responsible for its stability also supports the theoretical plausibility of a broader spectrum of deuteron-like dibaryon states. Nevertheless, despite decades of theoretical exploration and numerous model predictions\cite{Jaffe:1976yi,Mulders:1980vx,Balachandran:1983dj,Sakai:1999qm,Ikeda:2007nz,Bashkanov:2013cla,Shanahan:2011su,Clement:2016vnl}, no unambiguous experimental evidence for such states has been established to date~\cite{BaBar:2018hpv}.

Within this framework, baryonium states, comprising a baryon-antibaryon pair, constitute a distinctive subclass of hexaquark configurations. The interaction dynamics between a baryon and an antibaryon share important similarities with those governing baryon–baryon systems; however, baryon-antibaryon pairs can be produced with comparatively higher yields in high-energy collider experiments. Moreover, the presence of annihilation channels in baryon-antibaryon interactions can enhance the binding strength, potentially leading to more compact configurations. As such, the study of baryonium states offers a promising avenue for elucidating the dynamics of multiquark binding and may shed light on the underlying reasons for the apparent absence of experimentally confirmed stable dibaryon states.

The study of baryon-antibaryon systems dates back to the late 1940s, when Fermi and Yang proposed that $\pi$ mesons could be composite states formed from a nucleon-antinucleon pair~\cite{Fermi:1949voc}, a hypothesis that was later superseded by the quark model. At the dawn of the new millennium, the concept of heavy baryonium reemerged as a novel hadronic configuration, proposed as a possible explanation for the unusual properties of the $Y(4260)$ resonance~\cite{Qiao:2005av,Qiao:2007ce} and other experimentally observed charmonium-like states. In subsequent years, research on baryonium states expanded considerably, encompassing a wide range of theoretical approaches and phenomenological analyses~\cite{Chen:2011cta,Chen:2013sba,Wan:2019ake,Chen:2016ymy,Liu:2007tj,Wang:2021qmn,Wan:2022uie, Liu:2021gva,Wan:2021vny,Wan:2023epq,Wan:2025fyj}.

In the light hadron spectrum, the relatively small mass splittings between different states often make it challenging to distinguish exotic hadronic configurations from conventional hadrons, unless the former possess distinctive quantum numbers. Nevertheless, with access to large samples of $J/\psi$ events, the BESIII Collaboration has undertaken detailed studies of the energy region around 2.0 GeV~\cite{BESIII:2010gmv,BES:2003aic,BES:2005ega,BESIII:2010vwa,BESIII:2019wkp,BESIII:2016qzq,BESIII:2020vtu,BESIII:2017kqw,BESIII:2017hyw,BESIII:2019cuv,BESIII:2022tvj}, rekindling interest in light exotic hadronic states. 

In our earlier work, the masses of the light $N\bar{N}$, $\Lambda\bar{\Lambda}$, $\Sigma\bar{\Sigma}$, and $\Xi\bar{\Xi}$ baryonium states were systematically investigated~\cite{Wan:2021vny,Wan:2025fyj}. In particular, the mass of the $\Lambda\bar{\Lambda}$ state with quantum numbers $J^{PC}=1^{--}$ was predicted to be $(2.34\pm0.12)$ GeV. Subsequently, the BESIII Collaboration, through an analysis of the process $e^+e^-\to \Lambda\bar{\Lambda}\eta$, reported the observation of a $1^{--}$ structure in the $\Lambda\bar{\Lambda}$ invariant mass spectrum at $(2356\pm24)$ MeV with a decay width of $(304\pm82)$ MeV ~\cite{BESIII:2022tvj}. However, the substantial decay width of this enhancement complicates its unambiguous identification as a $\Lambda\bar{\Lambda}$ baryonium candidate, highlighting the experimental challenges inherent in the search for light baryonium states.

Given these difficulties, attention naturally turns to heavier baryon–antibaryon systems, where cleaner spectral signatures may be expected. In this context, the $\Omega\bar{\Omega}$ configuration presents a particularly compelling case. Composed entirely of strange quarks and antiquarks, it forms a pure $sss\bar{s}\bar{s}\bar{s}$ system, free from light-quark contamination and associated long-range pion-exchange effects. This unique quark composition allows for a more direct probe of short-range gluonic dynamics and multi-strange correlations in the baryon–antibaryon interaction. Furthermore, the relatively large mass of the $\Omega$ baryon, combined with the absence of strong decay channels below the dibaryon threshold, enhances the likelihood of observing narrow structures in experimental spectra. With high-luminosity facilities such as BESIII, Belle II, and LHCb capable of producing $\Omega\bar{\Omega}$ pairs in significant quantities, a dedicated search for such states is both timely and experimentally feasible.

In the present study, we perform a systematic investigation of the $\Omega\bar{\Omega}$ baryonium states with quantum numbers $J^{PC}=0^{-+}$, $1^{--}$, $0^{++}$, and $1^{++}$ within the framework of QCD sum rules (QCDSR). QCDSR offers a powerful nonperturbative tool that bridges the fundamental dynamics of QCD with hadronic phenomenology. Originally developed by Shifman, Vainshtein, and Zakharov~\cite{Shifman}, this methodology has been extensively and successfully applied to the study of both conventional and exotic hadrons~\cite{Albuquerque:2013ija,Wang:2013vex,Govaerts:1984hc,Reinders:1984sr,P.Col,Narison:1989aq,Tang:2021zti,Qiao:2014vva,Qiao:2015iea,Tang:2019nwv,Wan:2020oxt,Wan:2022xkx,Zhang:2022obn,Wan:2024dmi,Tang:2024zvf,Li:2024ctd,Zhao:2023imq,Yin:2021cbb,Yang:2020wkh,Wan:2024cpc,Wan:2024pet,Wan:2024ykm,Zhang:2024jvv,Tang:2024kmh,Tang:2016pcf,Tang:2015twt,Qiao:2013dda,Qiao:2013raa,Wan:2020fsk,Wan:2025xhf,Chen:2014vha,Azizi:2019xla,Wang:2017sto,Chung:1981wm,Wan:2025bdr}. In Sec.\ref{Formalism}, we provide a concise overview of the QCD sum rule framework and introduce the key formulas employed in our calculations. In Sec.\ref{Numerical}, the masses of the $\Omega\bar{\Omega}$ baryonium states with quantum numbers $J^{PC}=0^{-+}$, $1^{--}$, $0^{++}$, and $1^{++}$ are evaluated within the QCDSR approach. Section~\ref{decay} is devoted to an analysis of the possible decay channels of these $\Omega\bar{\Omega}$ baryonium states. Finally, a brief summary and concluding remarks are presented.

\section{Formalism}\label{Formalism}

Within the QCD sum rule formalism, the starting point is the analysis of two-point correlation functions constructed from suitably chosen interpolating currents. These two-point correlation encapsulate the nonperturbative dynamics of QCD and serve as a bridge between hadronic observables and the underlying quark–gluon degrees of freedom. They are defined as
\begin{eqnarray}
\Pi(q^2) &=& i \int d^4 x e^{i q \cdot x} \langle 0 | T \{ j (x),\;  j^\dagger (0) \} |0 \rangle\;,\label{twopoint1} \\ 
\Pi_{\mu\nu}(q^2) &=& i \int d^4 x e^{i q \cdot x} \langle 0 | T \{ j_\mu (x),\;  j_\nu^\dagger (0) \} |0 \rangle \; \label{twopoint2}
\end{eqnarray}
where, $j(x)$ and $j_\mu(x)$denote interpolating hadronic currents carrying total spin $J = 0$ and $1$, respectively, and $|0\rangle$ represents the QCD vacuum state. 
In the case of vector currents $j_\mu(x)$, Lorentz covariance dictates that the correlator admits the general decomposition
\begin{eqnarray}
\Pi_{\mu\nu}(q^2) &=&-\Big( g_{\mu \nu} - \frac{q_\mu q_\nu}{q^2}\Big) \Pi_1(q^2)+ \frac{q_\mu q_\nu}{q^2}\Pi_0(q^2)\;,
\end{eqnarray}
where $\Pi_1(q^2)$ and $\Pi_0(q^2)$ are invariant functions associated with intermediate hadronic states of the spin 1 and 0 mesons, respectively. 

In the QCD sum rules framework, interpolating currents are constructed to possess the same quantum numbers and partonic composition (quark and/or gluon content) as the hadronic states under investigation. Such currents serve as effective operators that couple directly to the physical states of interest, thereby enabling the extraction of their properties from first-principles QCD calculations. For the $\Omega\bar{\Omega}$ states with quantum numbers $ 0^{-+}$, $1^{--}$,  $0^{++}$, and $1^{++}$, the lowest-dimension local interpolating currents can be expressed as
\begin{eqnarray}\label{current_lambda}
j^{0^{-+}}(x)&=& i\,\bar{\eta}^\alpha_{\Omega}(x) \gamma_5 \eta^\alpha_{\Omega}(x) \;,\label{Ja0-+}\\
j^{1^{--}}_\mu(x)&=&\bar{\eta}^\alpha_{\Omega}(x) \gamma_\mu \eta^\alpha_{\Omega}(x)\;, \label{Ja1--}\\
j^{0^{++}}(x)&=& \bar{\eta}^\alpha_{\Omega}(x) \eta^\alpha_{\Omega}(x)\;,\label{Ja0++}\\
j^{1^{++}}_\mu(x)&=& i\,\bar{\eta}^\alpha_{\Omega}(x) \gamma_\mu\gamma_5 \eta^\alpha_{\Omega}(x) \;. \label{Ja1++}
\end{eqnarray}
Here, $\eta^\alpha_{\Omega}$ denotest the Dirac baryon fields carrying a Lorentz index $\alpha$, which interpolates the spin-$\frac{3}{2}$ $\Omega$ baryon. Following Ref. \cite{Chung:1981wm}, its explicit quark-level construction can be written as
\begin{eqnarray}\label{current_omega_b}
\eta^\alpha_{\Omega}(x)&=&i \epsilon_{a b c}[ s_a^{T}(x) C \gamma^\alpha s_b(x) ]s_c(x) \; ,
\end{eqnarray}
where $s_a(x)$ denotes the strange quark field with color index $a$, $C$ is the charge conjugation matrix, and $\epsilon_{a b c}$ is the totally antisymmetric tensor in color space. The superscript $T$ indicates transposition in Dirac space. This current configuration ensures a totally antisymmetric color wave function, consistent with the Pauli principle for baryonic states.

With the interpolating currents defined in Eqs. (\ref{Ja0-+})–(\ref{Ja1++}), the corresponding two-point correlation functions in Eqs. (\ref{twopoint1}) and (\ref{twopoint2}) can be straightforwardly constructed. In the QCD sum rule framework, these correlation functions admit two complementary representations: the Operator Product Expansion (OPE) representation, valid in the deep Euclidean region, and the phenomenological (hadronic) representation, which encodes the physical spectral information.

On the OPE side, short-distance (perturbative) and long-distance (nonperturbative) dynamics are systematically disentangled. The nonperturbative contributions are parameterized in terms of QCD vacuum condensates, such as quark, gluon, and mixed quark-gluon condensates, thereby incorporating the essential features of confinement and spontaneous chiral symmetry breaking into the formalism. In this representation, the invariant function $\Pi(q^2)$ can be expressed via a once-subtracted dispersion relation:
\begin{eqnarray}
\Pi(q^2) &=& \int_{s_{\text{min}}}^{\infty} ds \frac{\rho^{\text{OPE}}(s)}{s - q^2} \; , \label{OPE-hadron}
\end{eqnarray}
where $s_{min}$ denotes the kinematic threshold, typically given by the squared sum of the constituent (current) quark masses relevant to the hadronic channel under consideration~\cite{Albuquerque:2013ija}. The function $\rho^{OPE}(s) = \text{Im} [\Pi^{OPE}(s)] / \pi$ is the spectral density obtained from the OPE, incorporating contributions from local operators up to mass dimension 12. Explicitly, it can be decomposed as
\begin{eqnarray}
\rho^{OPE}(s) & = & \rho^{pert}(s) + \rho^{\langle \bar{s} s
\rangle}(s) +\rho^{\langle G^2 \rangle}(s) + \rho^{\langle \bar{s} G s \rangle}(s)
+ \rho^{\langle \bar{s} s \rangle^2}(s) +\rho^{\langle \bar{s} s \rangle\langle \bar{s} G s \rangle}(s) \nonumber\\
&+& \rho^{\langle \bar{s} s \rangle^3}(s)+ \rho^{\langle \bar{s} G s \rangle^2}(s) +\rho^{\langle \bar{s} s \rangle^2\langle \bar{s} G s \rangle}(s)+ \rho^{\langle \bar{s} s
\rangle^4}(s)\;. \label{rho-OPE}
\end{eqnarray}

In order to evaluate the spectral density on the OPE side, as given in Eq. (\ref{rho-OPE}), one requires the full coordinate-space propagator $S^q_{i j}(x)$ for a light quark $q\in {\{}u, d, s{\}}$. Incorporating both perturbative and nonperturbative effects, the light-quark propagator in the presence of QCD vacuum fields takes the form
\begin{eqnarray}
S^q_{j k}(x) \! \! & = & \! \! \frac{i \delta_{j k} x\!\!\!\slash}{2 \pi^2
x^4} - \frac{\delta_{jk} m_q}{4 \pi^2 x^2} - \frac{i t^a_{j k} G^a_{\alpha\beta}}{32 \pi^2 x^2}(\sigma^{\alpha \beta} x\!\!\!\slash
+ x\!\!\!\slash \sigma^{\alpha \beta}) - \frac{\delta_{jk}}{12} \langle\bar{q} q \rangle + \frac{i\delta_{j k}
x\!\!\!\slash}{48} m_q \langle \bar{q}q \rangle - \frac{\delta_{j k} x^2}{192} \langle g_s \bar{q} \sigma \cdot G q \rangle \nonumber \\ &+& \frac{i \delta_{jk} x^2 x\!\!\!\slash}{1152} m_q \langle g_s \bar{q} \sigma \cdot G q \rangle - \frac{t^a_{j k} \sigma_{\alpha \beta}}{192}
\langle g_s \bar{q} \sigma \cdot G q \rangle
+ \frac{i t^a_{jk}}{768} (\sigma_{\alpha \beta} x \!\!\!\slash + x\!\!\!\slash \sigma_{\alpha \beta}) m_q \langle
g_s \bar{q} \sigma \cdot G q \rangle \;.
\end{eqnarray}
The nonperturbative dynamics are manifest through the QCD vacuum condensates explicitly appearing in the above expression. The explicit form adopted here follows the standard treatment in the literature (see, e.g., Refs.~\cite{Wang:2013vex, Albuquerque:2013ija}), and constitutes the essential building block for the calculation of the OPE spectral density.

On the phenomenological side, the two-point correlation function is expressed in terms of hadronic degrees of freedom. In this representation, the interpolating current acts as an operator that creates and annihilates the hadronic state of interest from the QCD vacuum. By isolating the ground-state pole contribution from those arising from higher resonances and the continuum, the correlation function $\Pi(q^2)$ can be parameterized as
\begin{eqnarray}
\Pi^{\text{phen}}(q^2) &=& \frac{\lambda^2}{M^2 - q^2} + \frac{1}{\pi} \int_{s_0}^{\infty} ds \frac{\rho(s)}{s - q^2} \; , \label{hadron}
\end{eqnarray}
where $M$ denotes the physical mass of the hadronic ground state, $\lambda$ is the corresponding pole residue (current–hadron coupling constant), and $s_0$ is the continuum threshold parameter. The spectral density $\rho(s)$ in the second term encodes the contributions from excited states and the continuum, which are assumed to be dual to the perturbative QCD spectral density above $s_0$ in the quark–hadron duality ansatz.

Invoking the quark–hadron duality ansatz, the OPE representation and the phenomenological parameterization of the correlation function are equated, yielding
\begin{eqnarray}
\int_{s_{\text{min}}}^{\infty} ds \frac{\rho^{\text{OPE}}(s)}{s - q^2} &=& \frac{\lambda^2}{M^2 - q^2} + \frac{1}{\pi} \int_{s_0}^{\infty} ds \frac{\rho(s)}{s - q^2} \; , \label{mainSR}
\end{eqnarray}
which constitutes the basic QCD sum rule relation.

To improve the ground-state isolation, a Borel transformation is applied to Eq. (\ref{mainSR}). This procedure exponentially suppresses the contributions from higher resonances and continuum states, while simultaneously enhancing the pole term associated with the ground state. The hadron mass can then be extracted through
\begin{eqnarray}
M(s_0, M_B^2) &=& \sqrt{ - \frac{L_1(s_0, M_B^2)}{L_0(s_0, M_B^2)} } \; , \label{mass-Eq}
\end{eqnarray}
where $M_B^2$ denotes the Borel parameter. The auxiliary functions $L_0$ and $L_1$ are defined as
\begin{eqnarray}
L_0(s_0, M_B^2) &=& \int_{s_{\text{min}}}^{s_0} ds  \rho^{\text{OPE}}(s) e^{-s/M_B^2} \; , \label{L0}
\end{eqnarray}
\begin{eqnarray}
L_1(s_0, M_B^2) &=& \frac{\partial}{\partial (1/M_B^2)} L_0(s_0, M_B^2) \; .
\end{eqnarray}

\section{Numerical analysis}\label{Numerical}

In the numerical analysis, we adopt the widely accepted parameter values as reported in Refs.~\cite{Shifman,Albuquerque:2013ija,P.Col,Tang:2019nwv,ParticleDataGroup:2024cfk,Reinders:1984sr,Narison:1989aq}. Specifically, the strange-quark mass is taken as $m_s=(95\pm5)\; \text{MeV}$; the light-quark condensate is given by $\langle \bar{q} q \rangle = - (0.23 \pm 0.03)^3 \; \text{GeV}^3$; the strange-quark condensate is related via $\langle \bar{s} s \rangle=(0.8\pm0.1)\langle \bar{q} q \rangle$. For the mixed quark–gluon condensates, we use $\langle \bar{q} g_s \sigma \cdot G q \rangle = m_0^2 \langle\bar{q} q \rangle$ and $\langle \bar{s} g_s \sigma \cdot G s \rangle = m_0^2 \langle\bar{s} s \rangle$, with the mixed-condensate parameter $m_0^2 = (0.8 \pm 0.2) \; \text{GeV}^2$. The gluon condensate is taken as $\langle g_s^2 G^2 \rangle = (0.88\pm0.25) \; \text{GeV}^4$.

In addition to the aforementioned inputs, the QCD sum rule framework involves two auxiliary parameters: the continuum threshold $s_0$ and the Borel parameter $M_B^2$. Their values are fixed following the standard prescriptions in Refs.~\cite{Shifman,Reinders:1984sr,P.Col}, which are grounded in two widely accepted criteria. First, the convergence of the OPE is examined by comparing the magnitude of each higher-dimensional contribution to the total OPE result. The admissible range of $M_B^2$ is then chosen such that the truncated OPE series exhibits satisfactory convergence. Second, the pole contribution (PC) is evaluated to ensure the dominance of the ground-state signal. As emphasized in Refs.~\cite{Chen:2014vha,Azizi:2019xla,Wang:2017sto}, the spectral density for multiquark systems contains large powers of $s$, which tend to suppress the PC. In the case of hexaquark states, the PC is therefore required to be no less than $15\%$ of the total, thereby maintaining a sufficiently strong ground-state component. These two selection criteria can be formalized as follows:
\begin{eqnarray}
  R^{OPE}_{J^{PC}} = \left| \frac{L_{J^{PC},\;0}^{dim=12}(s_0, M_B^2)}{L_{J^{PC},\;0}(s_0, M_B^2)}\right|\, ,
\end{eqnarray}
\begin{eqnarray}
  R^{PC}_{J^{PC}} = \frac{L_{J^{PC},\;0}(s_0, M_B^2)}{L_{J^{PC},\;0}(\infty, M_B^2)} \; . \label{RatioPC}
\end{eqnarray}

The determination of an appropriate continuum threshold $s_0$ is carried out following the methodology outlined in Refs.\cite{Qiao:2013dda,Tang:2016pcf,Qiao:2013raa}. The guiding principle is to identify the value of $s_0$ that produces an optimal stability window in the mass curve of the baryonium state, within which the extracted mass exhibits minimal sensitivity to variations in the Borel parameter $M_B^2$. In practice, $s_0$ is varied by $0.1$ GeV to determine its lower and upper bounds, and the resulting range is used to estimate the associated uncertainty in $s_0$ \cite{Wan:2020oxt,Wan:2020fsk}. In Appendix B, we discuss in detail the significance of the continuum threshold parameter $\sqrt{s_0}$ within the QCD sum rule framework and its impact on the extracted mass values.

\begin{figure}
\includegraphics[width=6.8cm]{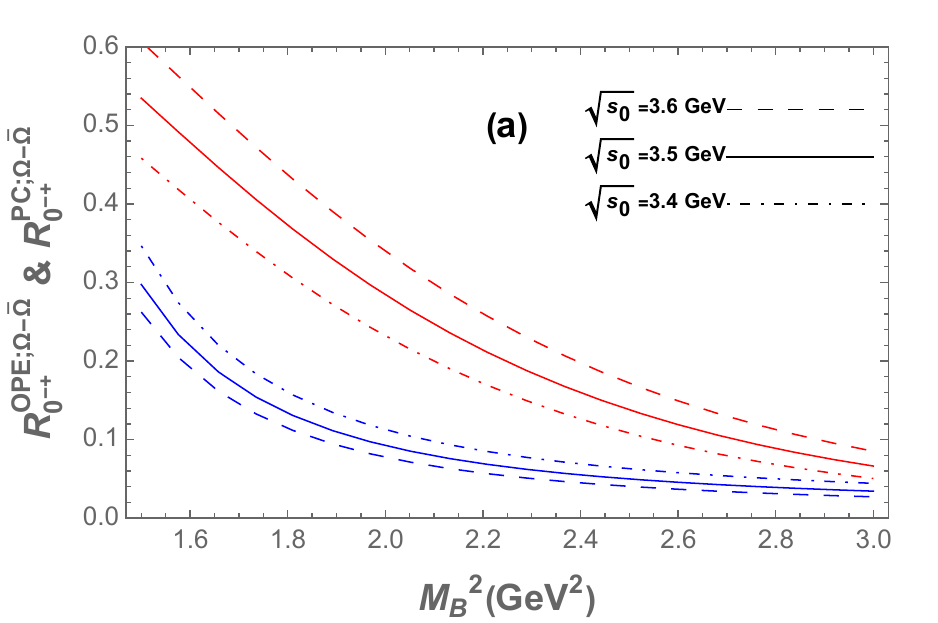}
\includegraphics[width=6.8cm]{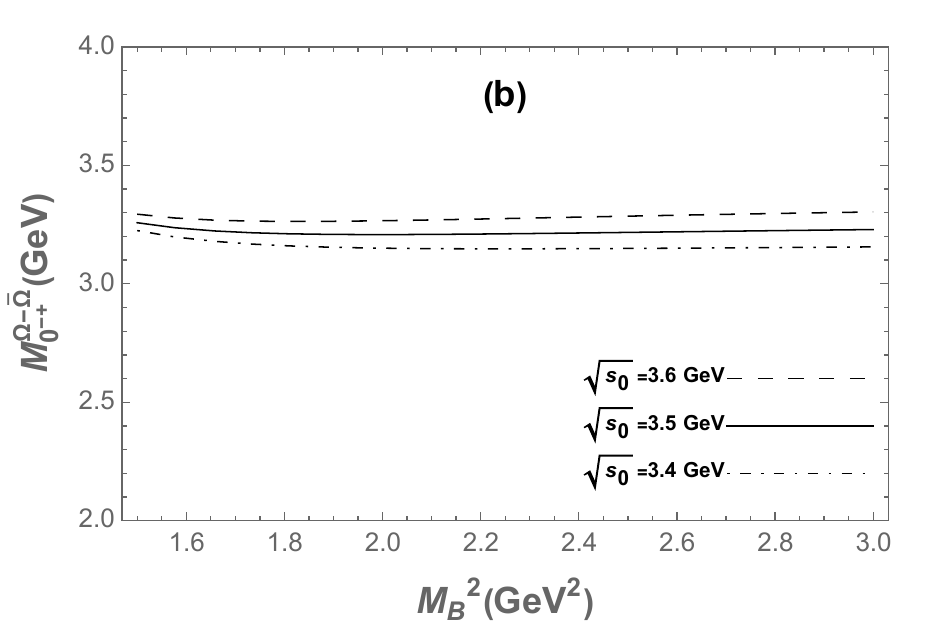}
\caption{ (a) The ratios of $R^{OPE}_{0^{-+}}$ and $R^{PC}_{0^{-+}}$ as functions of the Borel parameter $M_B^2$ for different values of $\sqrt{s_0}$, where blue lines represent $R^{OPE}_{0^{-+}}$ and red lines denote $R^{PC}_{0^{-+}}$. (b) The mass $m_{0^{-+}}$ as a function of the Borel parameter $M_B^2$ for different values of $\sqrt{s_0}$.} \label{fig0-+}
\end{figure}

With the above preparation the mass spectrum of the baryonium states can be numerically evaluated. For the $0^{-+}$ $\Omega\bar{\Omega}$ baryonium state in Eq.~(\ref{Ja0-+}), the ratios $R^{OPE}_{0^{-+}}$ and $R^{PC}_{0^{-+}}$ are presented as functions of Borel parameter $M_B^2$ in Fig. \ref{fig0-+}(a) with different values of $\sqrt{s_0}$, i.e., $3.4$, $3.5$, and $3.6$ GeV. The reliant relations of $m_{0^{-+}}$ on parameter $M_B^2$ are displayed in Fig. \ref{fig0-+}(b). The optimal Borel window is found in range $1.8 \le M_B^2 \le 2.6\; \text{GeV}^2$, and the mass $m_{0^{-+}}$ can then be obtained:
\begin{eqnarray}
m_{0^{-+}} &=& (3.22\pm 0.07)\; \text{GeV}.\label{m1}
\end{eqnarray}

\begin{figure}[h]
\includegraphics[width=6.8cm]{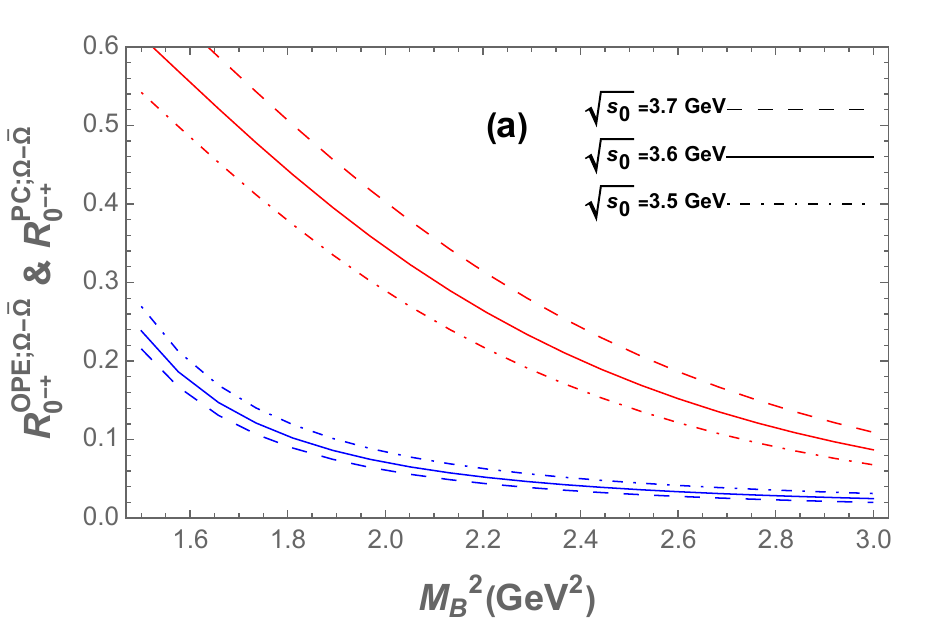}
\includegraphics[width=6.8cm]{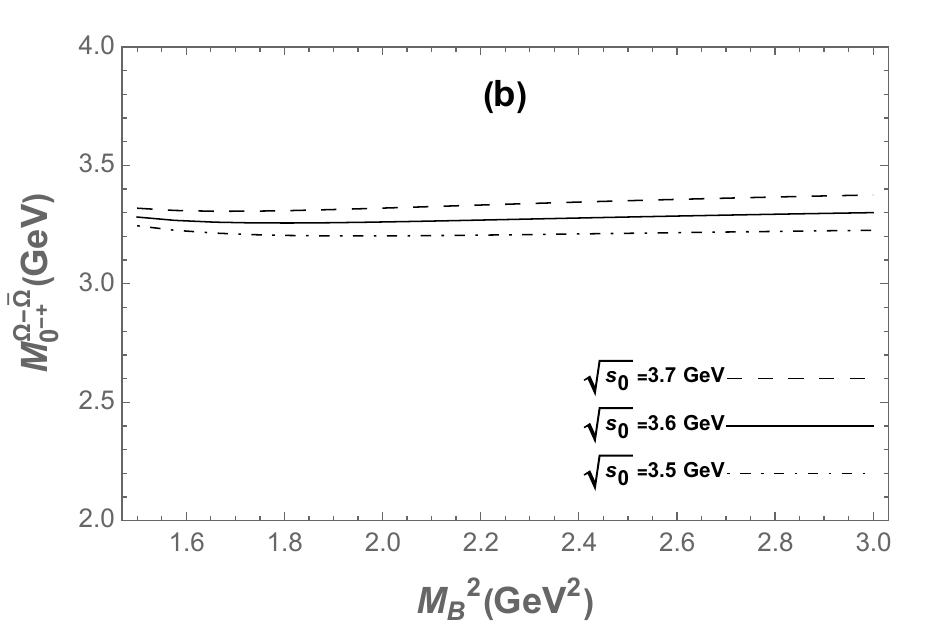}
\caption{(a) The ratios of $R^{OPE}_{1^{--}}$ and $R^{PC}_{1^{--}}$ as functions of the Borel parameter $M_B^2$ for different values of $\sqrt{s_0}$, where blue lines represent $R^{OPE}_{1^{--}}$ and red lines denote $R^{PC}_{1^{--}}$. (b) The mass $m_{1^{--}}$ as a function of the Borel parameter $M_B^2$ for different values of $\sqrt{s_0}$.} \label{fig1--}
\end{figure}

For the $1^{--}$ $\Omega\bar{\Omega}$ baryonium state defined in Eq.~(\ref{Ja1--}), the ratios $R^{OPE}_{1^{--}}$ and $R^{PC}_{1^{--}}$ are shown in Fig. \ref{fig1--}(a) as functions of Borel parameter $M_B^2$  with different values of $\sqrt{s_0}$, namely $3.5$, $3.6$, and $3.7$ GeV. The dependence of the extracted mass $m_{1^{--}}$ on $M_B^2$ is presented in Fig.~\ref{fig1--}(b). From this analysis, the optimal Borel window is determined to be $1.8 \le M_B^2 \le 2.6\; \text{GeV}^2$, within which the mass $m_{1^{--}}$ is obtained as:
\begin{eqnarray}
m_{1^{--}} &=& (3.28\pm 0.08)\; \text{GeV}.\label{m2}
\end{eqnarray}

\begin{figure}[h]
\includegraphics[width=6.8cm]{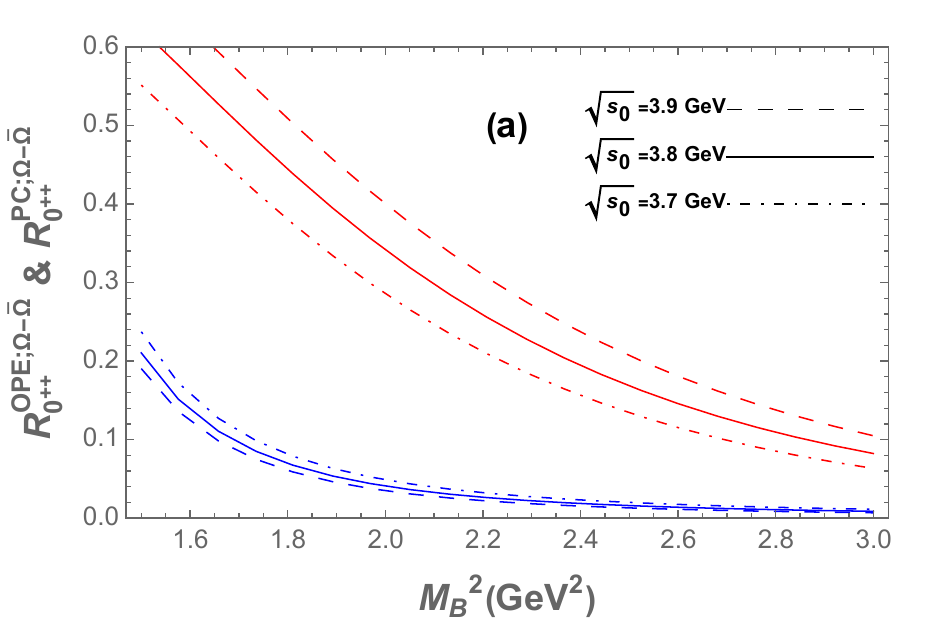}
\includegraphics[width=6.8cm]{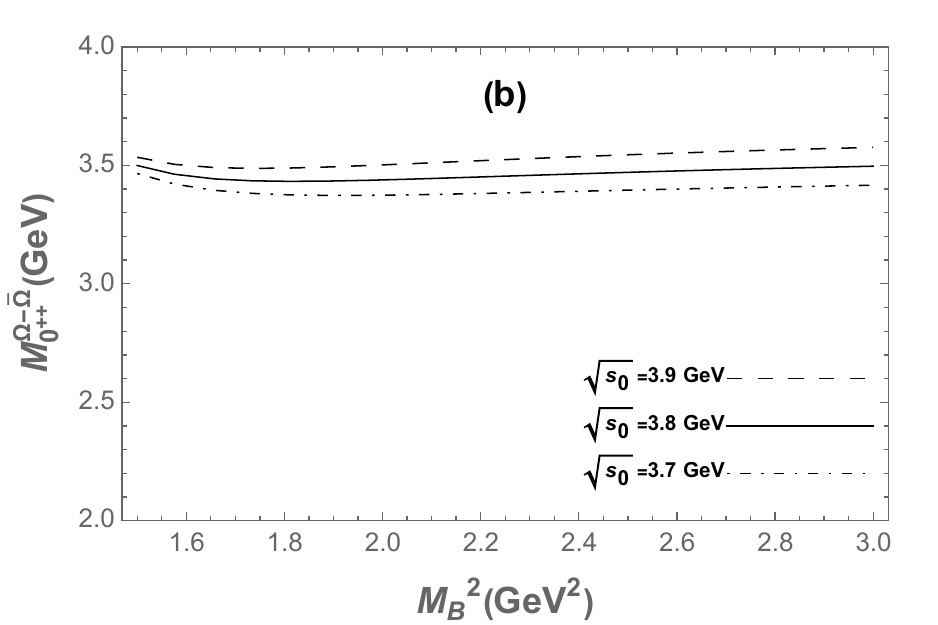}
\caption{(a) The ratios of $R^{OPE}_{0^{++}}$ and $R^{PC}_{0^{++}}$ as functions of the Borel parameter $M_B^2$ for different values of $\sqrt{s_0}$, where blue lines represent $R^{OPE}_{0^{++}}$ and red lines denote $R^{PC}_{0^{++}}$. (b) The mass $m_{0^{++}}$ as a function of the Borel parameter $M_B^2$ for different values of $\sqrt{s_0}$.} \label{fig0++}
\end{figure}

For the $0^{++}$ $\Omega\bar{\Omega}$ baryonium state defined in Eq.~(\ref{Ja0++}), the ratios $R^{OPE}_{0^{++}}$ and $R^{PC}_{0^{++}}$ are displayed in Fig. \ref{fig0++}(a) as functions of Borel parameter $M_B^2$ with different values of $\sqrt{s_0}$, namely $3.7$, $3.8$, and $3.9$ GeV. The dependence of the extracted mass $m_{0^{++}}$ on $M_B^2$ is shown in Fig.~\ref{fig0++}(b). Based on this analysis, the optimal Borel window is determined to be $1.7 \le M_B^2 \le 2.6\; \text{GeV}^2$, within which the mass $m_{0^{++}}$ is extracted as:
\begin{eqnarray}
m_{0^{++}} &=& (3.46\pm 0.09)\; \text{GeV}.\label{m3}
\end{eqnarray}

\begin{figure}[h]
\includegraphics[width=6.8cm]{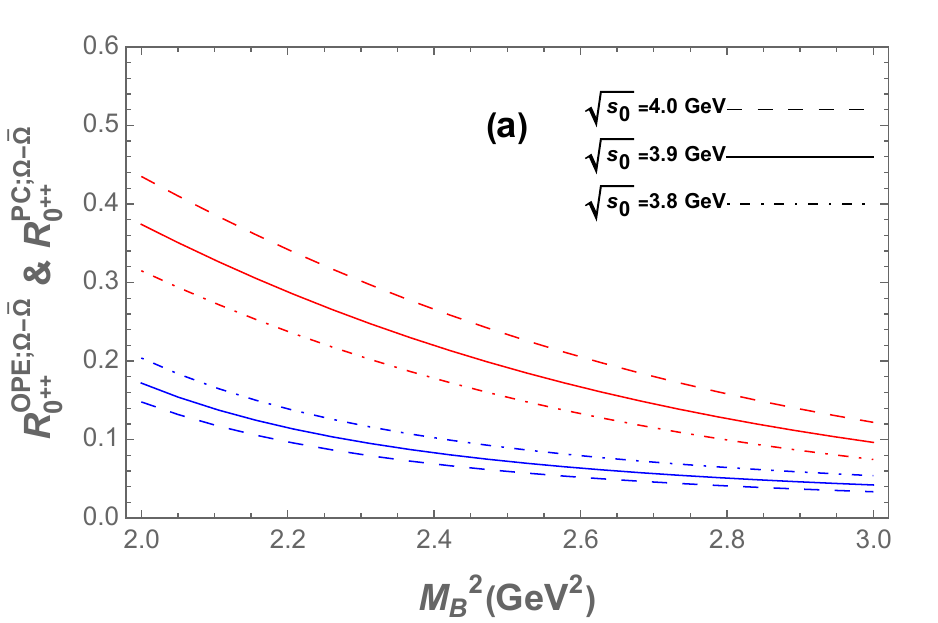}
\includegraphics[width=6.8cm]{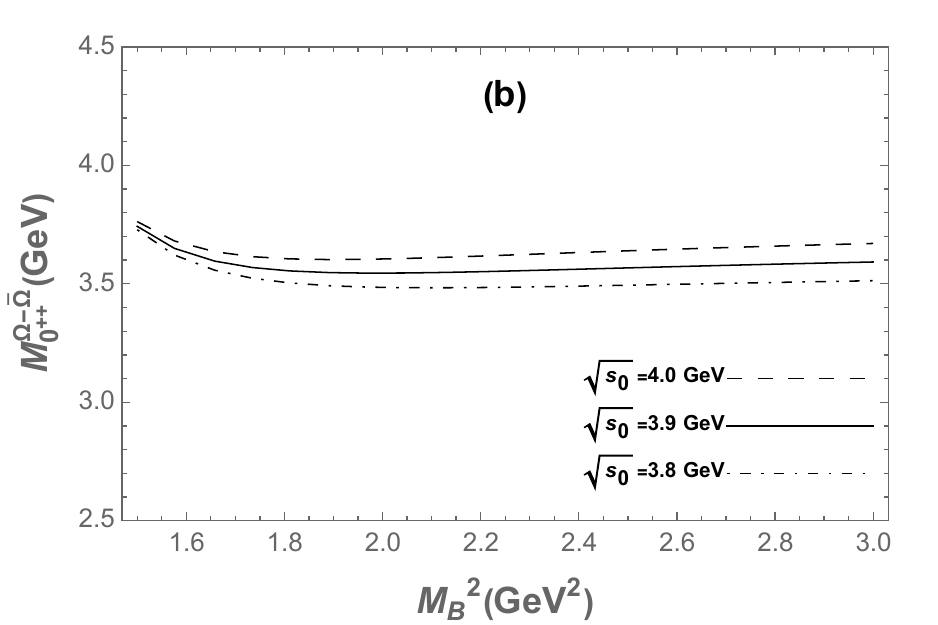}
\caption{(a) The ratios of $R^{OPE\;,B}_{0^{+-}}$ and $R^{PC\;,B}_{0^{+-}}$ as functions of the Borel parameter $M_B^2$ for different values of $\sqrt{s_0}$, where blue lines represent $R^{OPE\;,B}_{0^{+-}}$ and red lines denote $R^{PC\;,B}_{0^{+-}}$. (b) The mass $M^{B}_{0^{+-}}$ as a function of the Borel parameter $M_B^2$ for different values of $\sqrt{s_0}$.} \label{fig1++}
\end{figure}

For the $1^{++}$ $\Omega\bar{\Omega}$ baryonium state defined in Eq.~(\ref{Ja1++}), the ratios $R^{OPE}_{1^{++}}$ and $R^{PC}_{1^{++}}$ are illustrated in Fig. \ref{fig1++}(a) as functions of Borel parameter $M_B^2$ for several representative values of $\sqrt{s_0}$, specifically $3.8$, $3.9$, and $4.0$ GeV. The dependence of the extracted mass $m_{1^{++}}$ on $M_B^2$ is depicted in Fig.~\ref{fig1++}(b). From this analysis, the optimal Borel window is determined to lie within the range $2.2 \le M_B^2 \le 2.7\; \text{GeV}^2$, within which the mass $m_{1^{++}}$ is extracted as:
\begin{eqnarray}
m_{1^{++}} &=& (3.54\pm 0.11)\; \text{GeV}.\label{m4}
\end{eqnarray}

For clarity and ease of reference, the corresponding values of the Borel parameters, continuum thresholds, and the resulting mass predictions are summarized in Table~\ref{mass}.

The uncertainties in the present results primarily originate from the limited precision of the quark masses, vacuum condensates, and the continuum threshold parameter $\sqrt{s_0}$. Each parameter is varied independently within its accepted range, while keeping the others fixed at their central values. The resulting variation in the extracted mass is taken as the individual uncertainty associated with that parameter:
\begin{eqnarray}
\Delta m_i=|m(p_i^{max})-m(p_i^{min})|.
\end{eqnarray}
Here, $p_i$ denotes the parameters. The total uncertainty is estimated by summing individual contributions in quadrature:
\begin{eqnarray}
\Delta m_{totle}=\sqrt{\Sigma_i \Delta m_i^2}.
\end{eqnarray}
The numerical contributions from the continuum threshold, quark mass, and condensate values are summarized in Table~\ref{unc}. From the table, one can see that the dominant sources of uncertainty originate from the condensate values and the continuum threshold.

\begin{table}
\begin{center}
\renewcommand\tabcolsep{10pt}
\caption{The continuum thresholds, Borel parameters, and predicted masses of $\Omega\bar{\Omega}$ baryonium.}\label{mass}
\begin{tabular}{cccccc}\hline\hline
$J^{PC}$       & $\sqrt{s_0}\;(\text{GeV})$     &$M_B^2\;(\text{GeV}^2)$ &$M\;(\text{GeV})$       \\ \hline
 $0^{-+}$        & $3.5\pm0.1$                             &$1.8-2.6$                      &$3.22\pm0.07$         \\
 $1^{--}$         & $3.6\pm0.1$                             &$1.8-2.6$                      &$3.28\pm0.08$           \\
 $0^{++}$       & $3.8\pm0.1$                             &$1.7-2.6$                      &$3.46\pm0.09$         \\
 $1^{++}$       & $3.9\pm0.1$                             &$2.2-2.7$                      &$3.54\pm0.11$           \\
\hline                                                                                                                                                                      
 \hline
\end{tabular}
\end{center}
\end{table}

\begin{table}
\centering
\caption{Individual $\Delta m_i$ (GeV) contributions to the uncertainties of the extracted masses.}\label{unc}
\begin{tabular}{ccccc}
\hline\hline

 $J^{PC}$\;\;\;\;\; & from $s_0$ \;\;\;\;\;& from $m_s$ \;\;\;\;\;& from condensates \\
\hline

$0^{-+}$  & $\pm 0.05$ & $\pm 0.02$ & $\pm 0.04$ \\
$1^{--}$  & $\pm 0.05$ & $\pm 0.02$ & $\pm 0.06$ \\
$0^{++}$  & $\pm 0.07$ & $\pm 0.03$ & $\pm 0.05$ \\
$1^{++}$  & $\pm 0.08$ & $\pm 0.03$ & $\pm 0.07$ \\

\hline
\end{tabular}
\end{table}

\section{Decay analyses}\label{decay}

A definitive identification of these $\Omega\bar{\Omega}$ baryonium states can be achieved most directly through reconstruction from their decay products. Nevertheless, a thorough characterization of their intrinsic properties necessitates further systematic investigations. According to our calculations, the masses the $0^{-+}$ and $1^{--}$ states lie below the double-$\Omega$ threshold, whereas the masses of the $0^{++}$ and $1^{++}$ states exceed this threshold. Consequently, the sub-threshold states $0^{-+}$ and $1^{--}$ are kinematically forbidden from decaying into $\Omega\bar{\Omega}$ pairs; instead, their dominant decay modes are expected to be strong decays into three-meson final states, which are kinematically allowed. In contrast, the above-threshold states $0^{++}$ and $1^{++}$ can undergo strong decays into $\Omega\bar{\Omega}$ pairs. The detailed decay channels for these states are summarized below:
\begin{enumerate}
\item  For $J^{PC}=0^{-+}$ states, the primary decay channels are expected to be $\eta^{(\prime)}\eta^{(\prime)}\eta^{(\prime)}$, $\eta^{(\prime)}\omega\omega$, $\eta^{(\prime)}\omega\phi$, and $\eta^{(\prime)}\phi\phi$.
 \item  For $J^{PC}=1^{--}$ states, the dominant decay modes are anticipated to include $\eta^{(\prime)}\eta^{(\prime)}\omega$, $\eta^{(\prime)}\eta^{(\prime)}\phi$, $\omega\omega\omega$, $\omega\omega\phi$, $\omega\phi\phi$, and $\phi\phi\phi$.
 \item  For $J^{PC}=0^{++}$ states, the dominant decay channel is expected to be $\Omega\bar{\Omega}$, while the channels $\eta^{(\prime)}\eta^{(\prime)}f_0$ and $f_0f_0f_0$ are also kinematically favorable and likely to occur.
 \item  For $J^{PC}=1^{++}$ states, the principal decay channel is expected to be $\Omega\bar{\Omega}$, whereas the channels $\eta^{(\prime)}\eta^{(\prime)}f_1$ and $f_0 f_0 f_1$ are also kinematically allowed and likely to contribute.
\end{enumerate}

These decay channels are anticipated to be experimentally accessible at contemporary facilities such as BESIII, Belle II, and LHCb.

\section{Summary}

In summary, we have performed a comprehensive study of $\Omega\bar{\Omega}$ baryonium states with quantum numbers $J^{PC}=0^{-+}$, $1^{--}$, $0^{++}$, and $1^{++}$ within the framework of QCD sum rules. The numerical results, summarized in Table~\ref{mass}, indicate the possible existence of four baryonium states with masses $m_{0^{-+}}=(3.22\pm0.07)$ GeV, $m_{1^{--}}=(3.28\pm0.08)$ GeV, $m_{0^{++}}=(3.46\pm0.09)$ GeV, and $m_{1^{++}}=(3.54\pm0.11)$ GeV. The predicted masses of the $0^{-+}$ and $1^{--}$ states states lie below the corresponding double-$\Omega$, thresholds, suggesting that these states may exist as bound configurations. In contrast, the $0^{++}$ and $1^{++}$  states are found above their respective thresholds, indicating resonance-like characteristics. Potential decay channels of these $\Omega\bar{\Omega}$ baryonium states have been analyzed, and such processes are anticipated to be experimentally accessible at current facilities, including BESIII, Belle II, and LHCb.

It should be noted that the strong interaction in the $\Omega\bar{\Omega}$ system is expected to exhibit important similarities to that in the $\Omega\Omega$ system, in view of their closely related quark content and the underlying QCD dynamics. For the $\Omega\Omega$ ground state, lattice QCD calculations predict a binding energy of about 32 MeV in Ref.~\cite{Dhindsa:2025gae}, about 7 MeV in Ref.~\cite{Yamada:2014jra}, and about 1.6 MeV in Ref.~\cite{Gongyo:2017fjb}. These values show a slight discrepancy with our result, in which the binding energy of the $0^{-+}$ $\Omega\bar{\Omega}$ state is estimated to be approximately $(124.9 \pm 70.0)$ MeV. It is worth emphasizing, however, that the $\Omega\bar{\Omega}$ system corresponds to a baryonium configuration, where quark–antiquark annihilation effects may play a significant role, in contrast to the $\Omega\Omega$ dibaryon system. Indeed, within the framework of the chiral quark model and including the annihilation effect, the binding energy was found to be about 37–130 MeV in Ref.~\cite{Zhang:2006dy}, which is consistent with our result.


\vspace{.5cm} {\bf Acknowledgments} \vspace{.5cm}

This work was supported in part by the National Natural Science Foundation of China under Grants 12575106 and 12147214, and Specific Fund of Fundamental Scientific Research Operating Expenses for Undergraduate Universities in Liaoning Province under Grants No. LJ212410165019.


\begin{widetext}
\appendix
\section{The spectral densities}

The spectral densities, including contributions from the OPE, are presented below.

\begin{equation}\label{0-+}
\begin{aligned}
\rho(0^{-+})&=\frac{1}{23121100800\pi^{10}}\Biggl(642252800O3_s^{4}\pi^{8}
\left(10m_s^{2}+s\right)-1470s^{3}m_s^{2}O4 \left(20m_s^{2}+s\right) \\
&+40140800 O3_s^{2}O5_s\pi^{6}\left(111m_s^{3}+20s m_s\right)+3763200O5_s^{2}\pi^{4}
\left(66m_s^{4}+25m_s^{2}s+s^{2}\right) \\
& -3136m_s O3_s\pi^{2}s^{3}\left(2400m_s^{4}+120m_s^{2}s
+s^{2}\right)+15680 m_s s^{2} O5_s\pi^{2}\left(1680m_s^{4}
+160m_s^{2}s+3s^{2}\right) \\
& -40140800O3_s^{3}\pi^{6}\left(12m_s^{5}+69m_s^{3}s+4m_ss^{2}\right)\\
&-2508800 O3_s O5_s\pi^{4}
\left(24m_s^{6}+372m_s^{4}s+47m_s^{2}s^{2}+s^{3}\right) \\
& +s^{4}\left(70560m_s^{6}+4704m_s^{4}s+112m_s^{2}s^{2}+s^{3}\right)\\
&+125440O3_s^{2}\pi^{4}s
\left(720m_s^{6}+2400m_s^{4}s+155m_s^{2}s^{2}+2s^{3}\right)\Biggr).
\end{aligned}
\end{equation}

\begin{equation}\label{1--}
\begin{aligned}
\rho(1^{--}) &= \frac{1}{208089907200\pi^{10}} \Biggl(-11025 s^{3} m_s^{2} O4 \left (24m_s^{2} + s\right) +240844800003 s^{4} \pi^{8} \left(27m_s^{2} + 2s\right) \\
&+2408448000 O3_s^{2} O5_s \pi^{6} \left(198m_s^{3} + 25m_s s\right) +423360 m_s O5_s \pi^{2} s^{2} \left(580m_s^{4} + 48m_s^{2} s + s^{2}\right) \\
&-8064 m_s O3_s \pi^{2} s^{3} \left(8820m_s^{4} + 385m_s^{2} s + 4s^{2}\right) +42336000 O5_s^{2} \pi^{4}\left (576m_s^{4} + 160m_s^{2} s + 7s^{2}\right) \\
&-180633600O3_s^{3} \pi^{6} \left(18m_s^{5} + 160m_s^{3} s + 7m_s s^{2}\right) + s^{4}\left (635040m_s^{6} + 39312m_s^{4} s + 945m_s^{2} s^{2} + 8s^{3}\right) \\
&+188160O3_s^{2} \pi^{4} s\left (3600m_s^{6} + 15300m_s^{4} s + 792m_s^{2} s^{2} + 11s^{3}\right) \\
&-1128960O3_s O5_s \pi^{4} \left(360m_s^{6} + 8000m_s^{4} s + 785m_s^{2} s^{2} + 18s^{3}\right)\Biggr)
\end{aligned}
\end{equation}

\begin{equation}\label{0++}
\begin{aligned}
\rho(0^{++})&=\frac{1}{-23121100800\pi^{10}}\Biggl(-70560m_s^{6}s^{4}+s^{7}+9408m_s\pi^{2}s^{4}\left(-5O5_s+sO3_s\right)\\
&+11289600m_s^{5}\pi^{2}s^{2}\left(-3O5_s+sO3_s\right)+602112000m_s^{3}O3_s^{2}\pi^{6} \left(-13O5_s+9sO3_s\right)\\
&-1470m_s^{2}\left(6553600O3_s^{4}\pi^{8}-12800O5_s^{2}\pi^{4}s+25600O3_sO5_s\pi^{4}s^{2}-6400O3_s^{2}\pi^{4}s^{3}+O4s^{4}\right) \\
 &-29400m_s^{4}\left(11520O5_s^{2}\pi^{4}-46080O3_sO5_s\pi^{4}s+s^{2}\left(15360O3_s^{2}\pi^{4}-O4s\right)\right)
\Biggr)
\end{aligned}
\end{equation}

\begin{equation}\label{1++}
\begin{aligned}
\rho(1^{++}) &= \frac{1}{208089907200\pi^{10}} \Biggl(-11025 m_s^{2} O4 s^{3}\left (24m_s^{2} - s\right) +481689600O3_s^{4} \pi^{8} \left(165m_s^{2} + 2s\right) \\
&+240844800O3_s^{2} O5_s \pi^{6}\left (261m_s^{3} + 5m_s s\right) +42336000 O5_s^{2} \pi^{4} \left(672m_s^{4} + s^{2}\right) \\
&-180633600 m_s O3_s^{3} \pi^{6} \left(6m_s^{4} + 248m_s^{2} s + s^{2}\right) -40320 m_s O3_s \pi^{2} s^{3}\left (2436m_s^{4} + 7m_s^{2} s + 2s^{2}\right) \\
&+141120 m_s O5_s \pi^{2} s^{2} \left(2100m_s^{4} + 16m_s^{2} s + 3s^{2}\right) + s^{4} \left(635040m_s^{6} + 3024m_s^{4} s + 63m_s^{2} s^{2} - 8s^{3}\right) \\
&+188160O3_s^{2} \pi^{4} s\left (720m_s^{6} + 20700m_s^{4} s - 312m_s^{2} s^{2} + s^{3}\right) \\
&-1128960O3_s O5_s \pi^{4} \left(120m_s^{6} + 10240m_s^{4} s - 145m_s^{2} s^{2} + 2s^{3}\right)\Biggr)
\end{aligned}
\end{equation}

Here, $O3_s$, $O4$, and $O5_s$ are shorthand for $\langle \bar{s} s\rangle$, $\langle G^2 \rangle$, and $\langle \bar{s} G s \rangle$, respectively.

\section{Role of the Continuum Threshold Parameter $\sqrt{s_0}$ in QCD Sum Rule Analysis}

To examine the sensitivity of the extracted mass to the continuum threshold parameter, we vary $\sqrt{s_0}$ within a reasonable range around its central value. As shown in Fig.~\ref{s00-+}, although the mass exhibits a slight dependence on $\sqrt{s_0}$, a stable platform can still be observed in the chosen Borel window. This behavior indicates that our results are not overly sensitive to the choice of $\sqrt{s_0}$, and the corresponding uncertainty has been properly taken into account in the final error estimation.

\begin{figure}
\includegraphics[width=6.8cm]{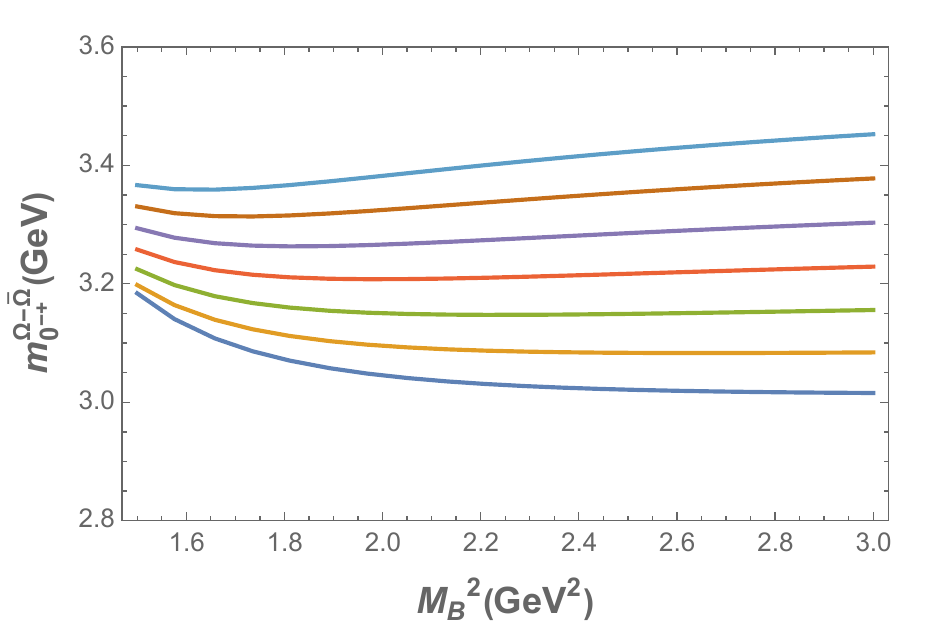}
\caption{ In the figure, the curves from bottom to top correspond to $\sqrt{s_0}$ values of 3.2, 3.3, 3.4, 3.5, 3.6, 3.7, and 3.8 GeV, respectively..} \label{s00-+}
\end{figure}

For the $0^{-+}$ state, as shown in Fig.~\ref{s00-+}, the curve is the flattest when $\sqrt{s_0} = 3.5$ GeV, indicating the weakest dependence of the mass on the Borel parameter. When $\sqrt{s_0} = 3.3$ GeV, the curve shows a noticeable downward trend, while for $\sqrt{s_0} = 3.7$ GeV, the curve exhibits a clear upward trend. Therefore, we choose $\sqrt{s_0} = 3.5$ GeV.

A similar analysis has been performed for the other states, with the corresponding figures shown in Fig.~\ref{s01--}, Fig.~\ref{s00++}, and Fig.~\ref{s01++}, which correspond to the $1^{--}$, $0^{++}$, and $1^{++}$ states, respectively.

\begin{figure}
\includegraphics[width=6.8cm]{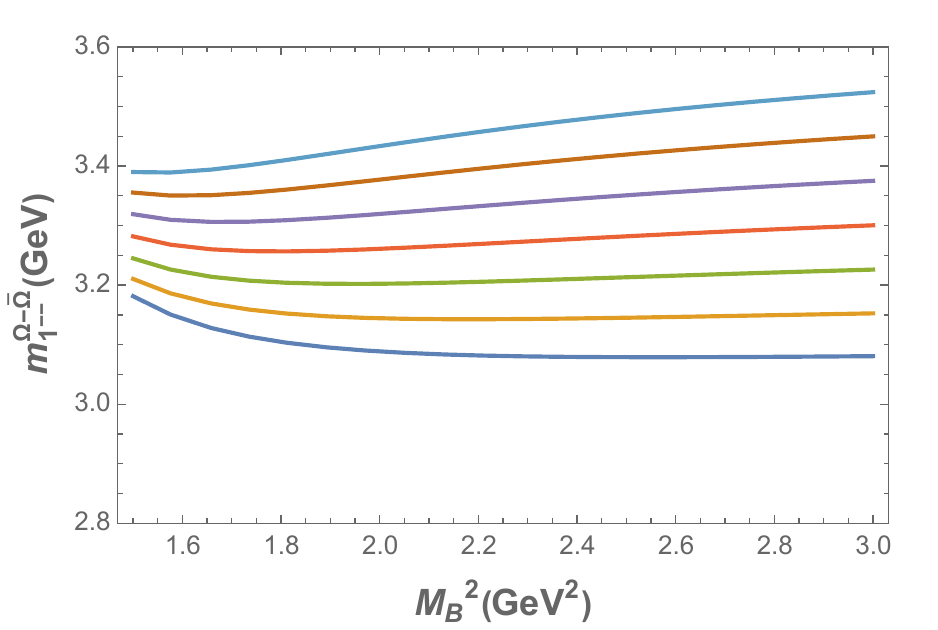}
\caption{ In the figure, the curves from bottom to top correspond to $\sqrt{s_0}$ values of 3.3, 3.4, 3.5, 3.6, 3.7, 3.8, and 3.9 GeV, respectively..} \label{s01--}
\end{figure}

\begin{figure}
\includegraphics[width=6.8cm]{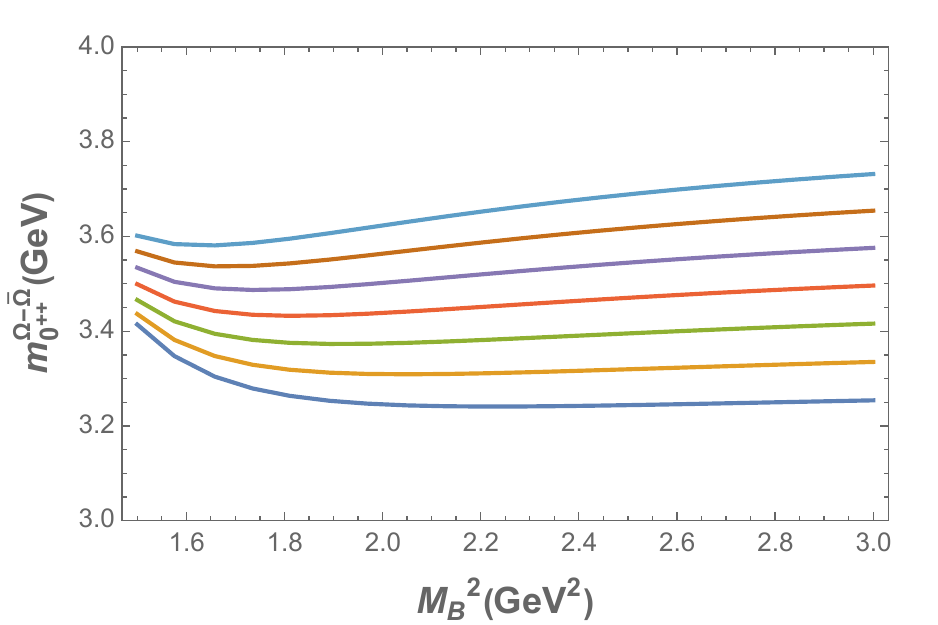}
\caption{ In the figure, the curves from bottom to top correspond to $\sqrt{s_0}$ values of 3.5, 3.6, 3.7, 3.8, 3.9, 4.0, and 4.1 GeV, respectively..} \label{s00++}
\end{figure}

\begin{figure}
\includegraphics[width=6.8cm]{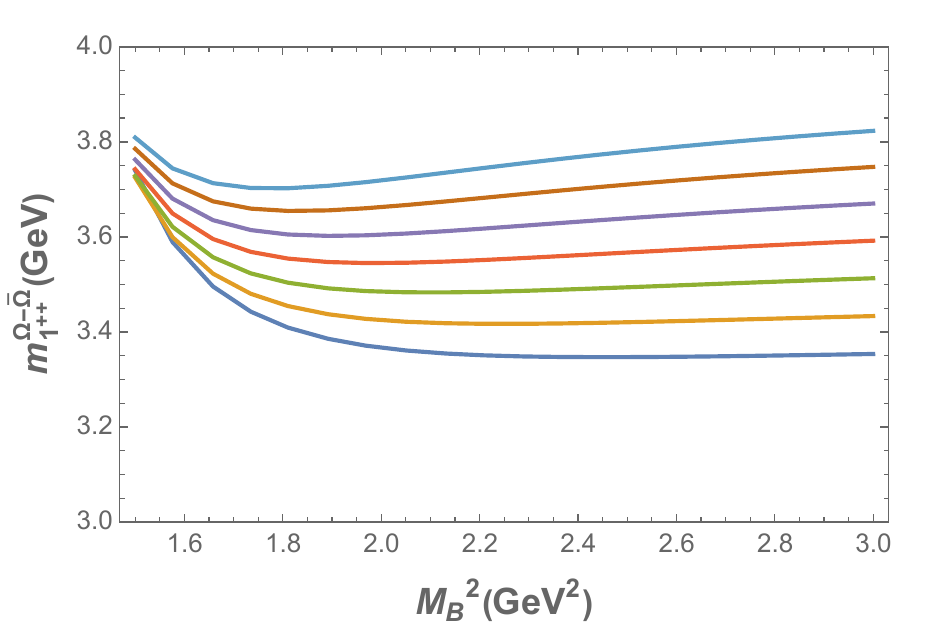}
\caption{ In the figure, the curves from bottom to top correspond to $\sqrt{s_0}$ values of 3.6, 3.7, 3.8, 3.9, 4.0, 4.1, and 4.2 GeV, respectively..} \label{s01++}
\end{figure}

\end{widetext}

\end{document}